\documentclass[11pt,twocolumn,amsmath,amssymb,showkeys,natbib,rmp]{revtex4}

\usepackage{graphicx}
\usepackage{dcolumn}
\usepackage{bm}

\begin{document}


\title{Noise effects in polymer dynamics}

\author{N. Pizzolato$^{a,}$\footnote{e-mail address: npizzolato@gip.dft.unipa.it},
A. Fiasconaro$^{ab}$, B. Spagnolo$^a$} \affiliation{$^a$Dipartimento
di Fisica e Tecnologie Relative, Universit\`a di Palermo and\\
CNISM-INFM, Unit\`a di Palermo, Group of Interdisciplinary
Physics\footnote{URL: http://gip.dft.unipa.it},\\
Viale delle Scienze, edificio 18, I-90128 Palermo, Italy\\
$^b$Mark Kac Complex Systems Research Center, Institute of Physics,
Jagellonian University, Reymonta 4, 30-059 Krak$\acute{o}$w, Poland}


\begin{abstract}
The study of the noise induced effects on the dynamics of a chain
molecule crossing a potential barrier, in the presence of a
metastable state, is presented. A two-dimensional stochastic
version of the Rouse model for a flexible polymer has been adopted
to mimic the molecular dynamics and to take into account the
interactions between adjacent monomers. We obtain a nonmonotonic
behavior of the mean first passage time and its standard
deviation, of the polymer centre of inertia, with the noise
intensity. These findings reveal a noise induced effect on the
mean crossing time. The role of the polymer length is also
investigated.
\end{abstract}

\maketitle

\section{\label{intro}Introduction}
Many important biological processes are governed by the transport of
molecules across membranes. In cellular systems proteins translocate
from the cytosol into the endoplasmatic reticulum [Hu \emph{et al.},
2005; Chiam \emph{et al.}, 2006], or into mithocondria or
chloroplasts [Frappat \emph{et al.}, 2003]; RNA molecules are
transported across a nuclear membrane after their synthesis [Schatz
\& Dobberstein, 1996]; in infections from bacteria resistent to
antibiotics, alternative therapy uses bacteriophages which push a
long chain of DNA molecules through the pores of a membrane [Adhya
\& Merril, 2006]. The study of the dynamics of translocation of
individual polymers across nanometer-scale pores is essential to
help the understanding of how biological systems work and it also
represents a powerful instrument to investigate the structural
characteristics and conformational changes of chain molecules, their
interactions with the walls of the pore and specific folding or
unfolding processes related to the crossing event. A fundamental
work on polymer translocation was carried out by Kasianowicz
\emph{et al.}~[1996]. In their experiments, single-stranded DNA
molecules are forced by an applied electric field to move through a
2.6-nm diameter ion channel in a voltage-biased lipid bilayer;
during the passage across the pore, the polymer significantly
reduces the single-channel ion current. By measuring the duration
time of each current drop - blockade -, they have demonstrated that
the crossing time linearly increases with the polymer length. By
using a similar experimental setup, Meller \emph{et al.}~[2002] have
also shown that the time-length relationship depends on the number
of adenines in the polymer, concluding that the translocation time
is "heavily" influenced by specific interactions between the DNA and
the nanopore walls. They also suggest stronger interactions at lower
temperatures because of the longer measured translocation times
[Meller \emph{et al.}, 2000]. On the other hand, very interesting
experimental results by Han \emph{et al.}~[1999] show longer
crossing times for shorter DNA molecules, suggesting the existence
of a quasi-equilibrium state of the polymer after being trapped for
a time. Theoretical studies have demonstrated that the translocation
time of a long N-segments chain molecule, diffusing across a
membrane, scales as N$^3$ if there is no free energy bias between
the two sides, or as N$^2$ in case of adsorption [Sung \& Park,
1996; Park \& Sung, 1998]. Strong interactions between the polymer
and the pore modify the dynamics of the portion of the chain inside
the channel, making linear the relationship between the crossing
time and N [Lubensky \& Nelson, 1999]. In the presence of an
externally applied potential, the linear dependence of the crossing
time on the length of the polymer is confirmed only for the case of
an hairpin crossing mechanism of translocation [Schatz \&
Dobberstein, 1996; Sebastian \& Paul, 2000]. The dependence of the
crossing time on the length of the polymer chain, as reported in
different experimental works, may be influenced by the different
geometrical and physical characteristics of the pore-channel device,
the adopted polynucleotides, the intensity of the driving electric
fields. We are convinced that, in same cases, a special condition
for the polymer travelling across a nanometer-scale channel can be
realized to make it temporarily trapped into a metastable state, as
already suggested by Han \emph{et al.}~[1999]. For this reason, we
decided to start a theoretical study of the dynamics of polymers in
a state out of equilibrium under the influence of a noisy
environment. In this paper, we firstly describe the model and the
mathematical framework adopted to simulate the polymer dynamics and,
finally, we comment our results.

\section{\label{model} The model}
\begin{figure}[htbp]
\begin{center}
\vskip+0.3cm
\includegraphics[width=7.0cm]{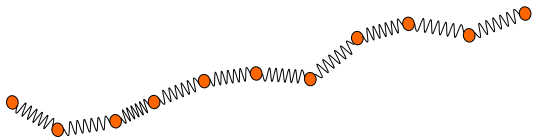}
\end{center}
\vskip-0.6cm \caption{\small Schematic representation of the polymer
chain.}
 \label{fig1}
\end{figure}

\begin{figure}[htbp]
\begin{center}
\vskip-1.6cm
\includegraphics[width=7.0cm,height=8.0cm]{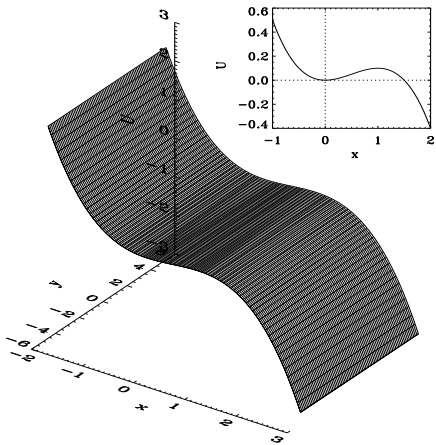}
\end{center}
\vskip-0.6cm \caption{\small Potential surface of the
two-dimensional barrier. Inset: projection of the potential
surface on the plane $y=0$.}
 \label{fig2}
\vskip-0.5cm
\end{figure}

Stochastic modelling of polymer dynamics starts with the choice of a
model for the chain molecule. In the simplest model of a polymer,
monomers are freely-jointed rigid rods of a fixed length (Kuhn
length) and any kind of interactions among them is neglected. For
our studies, we have adopted the Rouse model for a flexible polymer
[Rouse, 1953; Doi \& Edwards, 1986]. This model describes the
molecule as a chain of N monomers connected by harmonic springs
(Fig.~\ref{fig1}).
\\
At this stage, we have neglected the excluded-volume effect and any
hydrodynamic interaction between monomers and solvent, as modelled
by the Zimm theory [Zimm, 1956]. In a recent experiment, a
fluorescent correlation spectroscopy technique has been applied to
monitor the stochastic motion of individual monomers within isolated
DNA molecules [Shusterman \emph{et al.}, 2004]. By comparing the
observed time dependence of the mean-squared diplacement of the
fluorescent monomer with that predicted by the available models,
Shusterman and coworkers assert that double stranded DNA (dsDNA)
behaves like a semiflexible coil following the Rouse model, while in
the case of single stranded DNA (ssDNA) the monomers show a
Zimm-type kinetics. A similar analysis of the short-time dynamics
demonstrates that intramolecular hydrodynamic interactions in dsDNA
chains cannot be ruled out [Tothova \emph{et al.}, 2005; Petrov
\emph{et al.}, 2006], suggesting the application of a mixed
Rouse-Zimm model [Lisy \emph{et al.}, 2005]. We have decided to
start our investigation by using a relatively simple bead-spring
model in order to avoid any secondary effect of metastability
induced by the combination of an excessive number of parameters.\\
The translocation of a long molecule through a narrow pore is caused
by a forced process of unfolding by which the coil is stretched
across the channel, returning in its native folded state again at
the exit only. Within the pore region, the free energy per segment
of the polymer is higher, representing a barrier for the motion of
the molecule [Sung \& Park, 1996]. In the present work, we study the
effect of the temperature fluctuations on the dynamics of a chain
polymer escaping from a metastable state through a potential barrier
in a two-dimensional domain. The polymer motion is modeled as a
stochastic process of diffusion in the presence of a
time-independent potential having a profile described by the
following equation:
\begin{equation}
\label{eq1} U(x)=ax^2-bx^3,
\end{equation}
with parameters $a$=0.3 and $b$=0.2, as already adopted by
Fiasconaro \emph{et al.}~[2003] for the unidimensional
single-particle case. The corresponding 2D-surface is shown in
Fig.~\ref{fig2}. The drift of the i-th monomer of the chain is
described by the following coupled Langevin equations [Gardiner,
1993]:
\begin{align}
&\frac{\partial{x}}{\partial{t}}=-\frac{\partial{U(x)}}{\partial{x}}+F_{int}(x)+\sqrt{D}\xi_x(t),
\label{eq2}\\
&\frac{\partial{y}}{\partial{t}}=F_{int}(y)+\sqrt{D}\xi_y(t),
\label{eq3}
\end{align}
where $F_{int}(x)$ and $F_{int}(y)$ are the $x$ and $y$ components,
respectively, of the harmonic forces between adjacent monomers and
$\xi_x(t)$ and $\xi_y(t)$ are white Gaussian noise modeling the
temperature fluctuations, with the usual statistical properties,
namely $\langle\xi_i(t)\rangle=0$ and $ \langle
\xi_i(t)\xi_j(t+\tau)\rangle =\delta_{ij}(\tau)$ for $(i,j=x,y)$. We
have numerically solved the above system of equations by performing
a set of $10^5$ simulations for 22 different values of the noise
intensity $D$ and 5 different polymer length configurations. In this
study, we put the starting distance $L_0$ between two adjacent
monomers of the molecule chain equal to the rest length of the ideal
spring connecting them. We have chosen $L_0=0.01$ and the harmonic
constant equal to 1.0 (in arbitrary unit). The number $N$ of
monomers changes between 2 and 10. The initial position is set to
$x_0$=1.3 for all monomers; every simulation stops when the $x$
coordinate of the center of mass of the chain reaches the final
position at $x_f$=2.2. Then, the Mean First Passage Time (MFPT) and
its standard deviation are calculated.
\\
\section{\label{Results}Results}
\begin{figure*}[htbp]
\begin{center}
\hspace{-0.4cm}
\includegraphics[width=8cm,height=7cm]{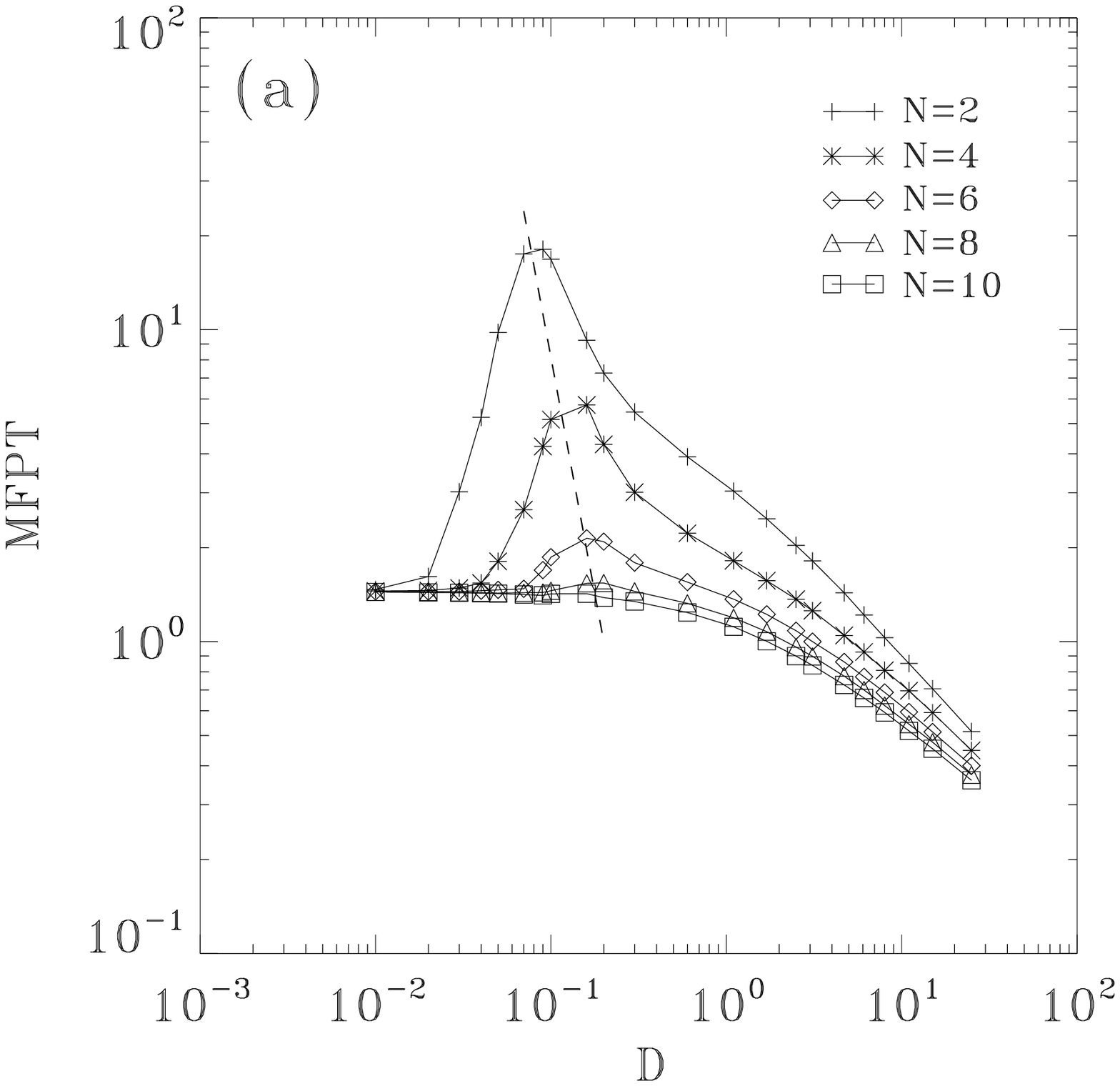}
\hspace{+0.5cm}
\includegraphics[width=8cm,height=7cm]{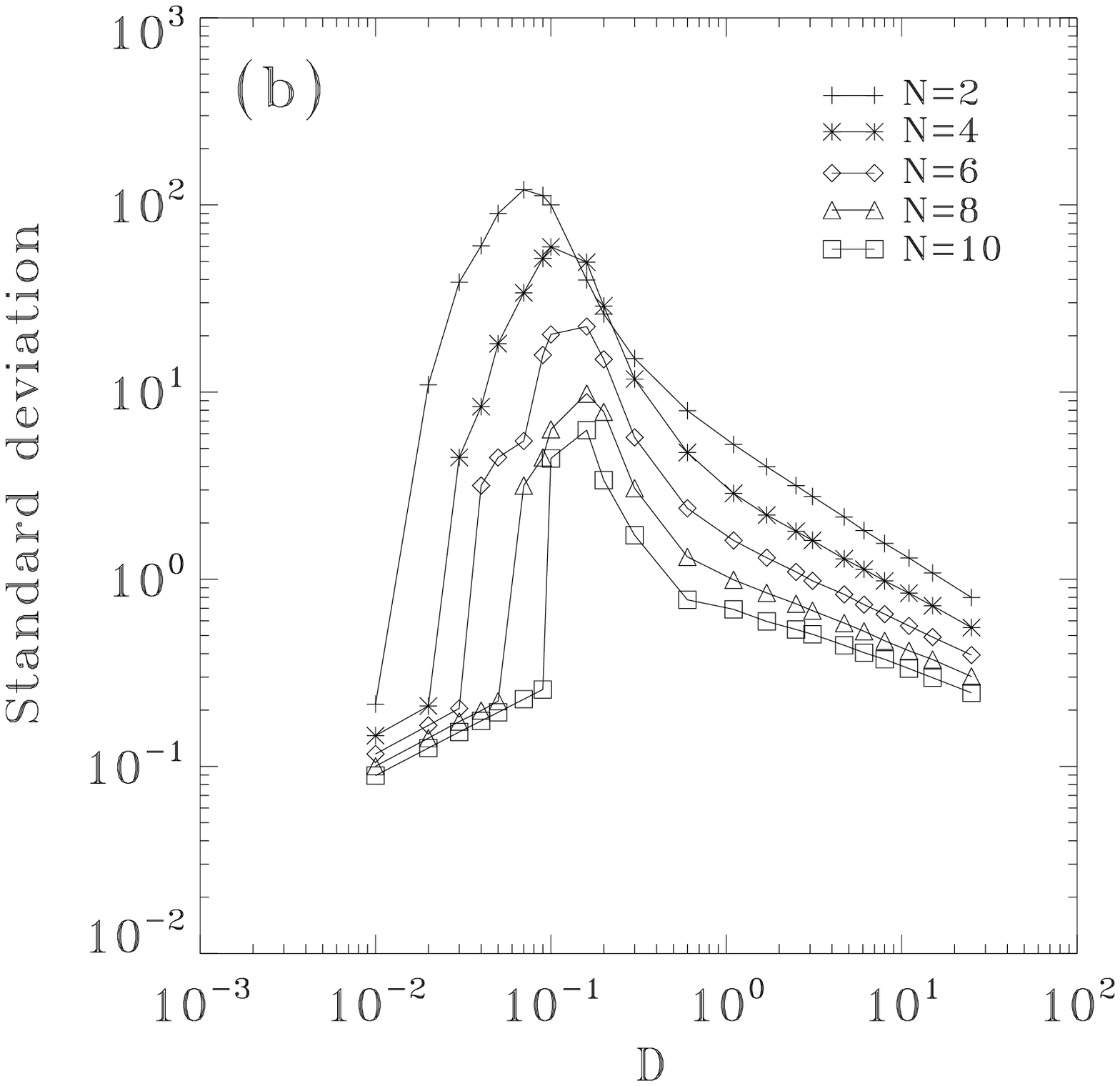}
\end{center}
\vskip-0.6cm \caption{\small (a): Mean first passage time vs.~noise
intensity for five different numbers N of monomers in the polymer.
The starting distance between adjacent monomers coincides with the
rest distance $L_0=0.01$. The initial condition $x_0=1.3$ is the
same for all the monomers. The dashed line indicate the shift of the
maximum of the MFPT. (b): Standard deviation of the passage time vs.
noise intensity for different polymer configurations.}
\label{fig34}
\vskip+0.0cm
\end{figure*}

\begin{figure}[htbp]
\begin{center}
\includegraphics[width=7.5cm]{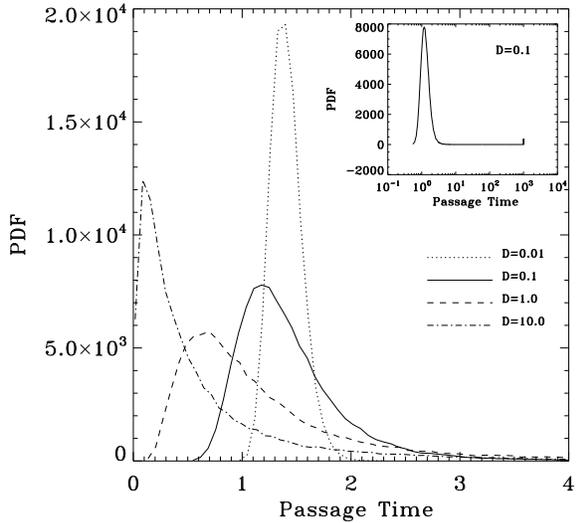}
\end{center}
\vskip-0.6cm \caption{\small Probability density function (PDF) of
the passage time of the polymer center of mass with N=4 at four
different values of the noise intensity. The inserted plot shows the
tail at high passage times of the PDF for $D=0.1$.}
 \label{fig5}
\end{figure}

First results of this study suggest the presence of the noise
enhanced stability (NES) effect [Mantegna \& Spagnolo, 1996; Agudov
\& Spagnolo, 2001; Fiasconaro \emph{et al.}, 2003; Fiasconaro
\emph{et al.}, 2005] for a short polymer escaping from a potential
barrier. In the left panel of Fig.~\ref{fig34} we show how the MFPT
changes with the noise intensity at different polymer lengths.
Assuming identical unstable initial conditions we find that: (i) the
NES effect is more relevant for shorter chains (lower number of
monomers); (ii) the maximum of the MFPT rapidly decreases at
increasing polymer length and its position shifts towards higher
levels of noise.
\\
The dynamics of shorter chains is much more similar to that of a
single particle escaping from the potential well. As a consequence
shorter chains show more pronounced NES effect, with higher maximum
values of the MFPT. Moreover, the shift of the position of this
maximum is directly connected with the polymer length. In fact,
longer polymers need higher noise intensity values to move the
centre of mass back into the metastable state, starting from initial
unstable states. Longer chain polymers experience more than shorter
ones the effect of the potential slope in the unstable region. It is
important to note that the behavior of MFPT at very low noise
intensities ($D\leq 5\cdot10^{-2}$, in Fig.~\ref{fig34}a) should
have divergent trend because the initial unstable states used in our
simulations are in the divergent dynamical regime (detailed analysis
on the divergent dynamical regime can be found in Fiasconaro
\emph{et al.}~[2005]). To find this behavior we need very long
computer simulation times, which become greater as the polymer
length increases. In fact the probability that the entire sequence
of $N$ monomers experience the potential well is much lower than the
single particle, and so more difficult to see in a finite time
computer simulation.\\ The plot of the standard deviation of the
first passage time as a function of the noise intensity $D$ is shown
in Fig.~\ref{fig34}b. An increasing of the standard deviation means
a spread of the passage time distribution and this represents a
direct evidence of the noise-induced capture effect of the barrier.
In Fig.~\ref{fig5} we report the probability density function (PDF)
of the passage time of the polymer center of mass for the case of
N=4 and four values of the noise intensity, namely $D=0.01$, 0.1,
1.0 and 10.0. The PDFs obtained with lower noise intensities are
characterized by a strong asymmetry and high long tails. In this
figure we can clearly see that, while the most probable value
follows a monotonic decreasing trend with increasing noise
intensity, the MFPT is strongly affected by the rare events of
trapped particles giving the long tail of the distribution at high
values of the passage time (see inset in Fig.~\ref{fig5}). The shape
of the tail significantly influences the mean crossing time by
increasing its value with respect to the mode. For the case of N=10,
the plot of MFPT vs.~$D$ does not show any evidence of a NES effect,
but not the standard deviation, which shows a non-monotonic
behavior.\\
\begin{figure*}[htbp]
\begin{center}
\hspace{-0.4cm}
\includegraphics[width=8cm,height=7cm]{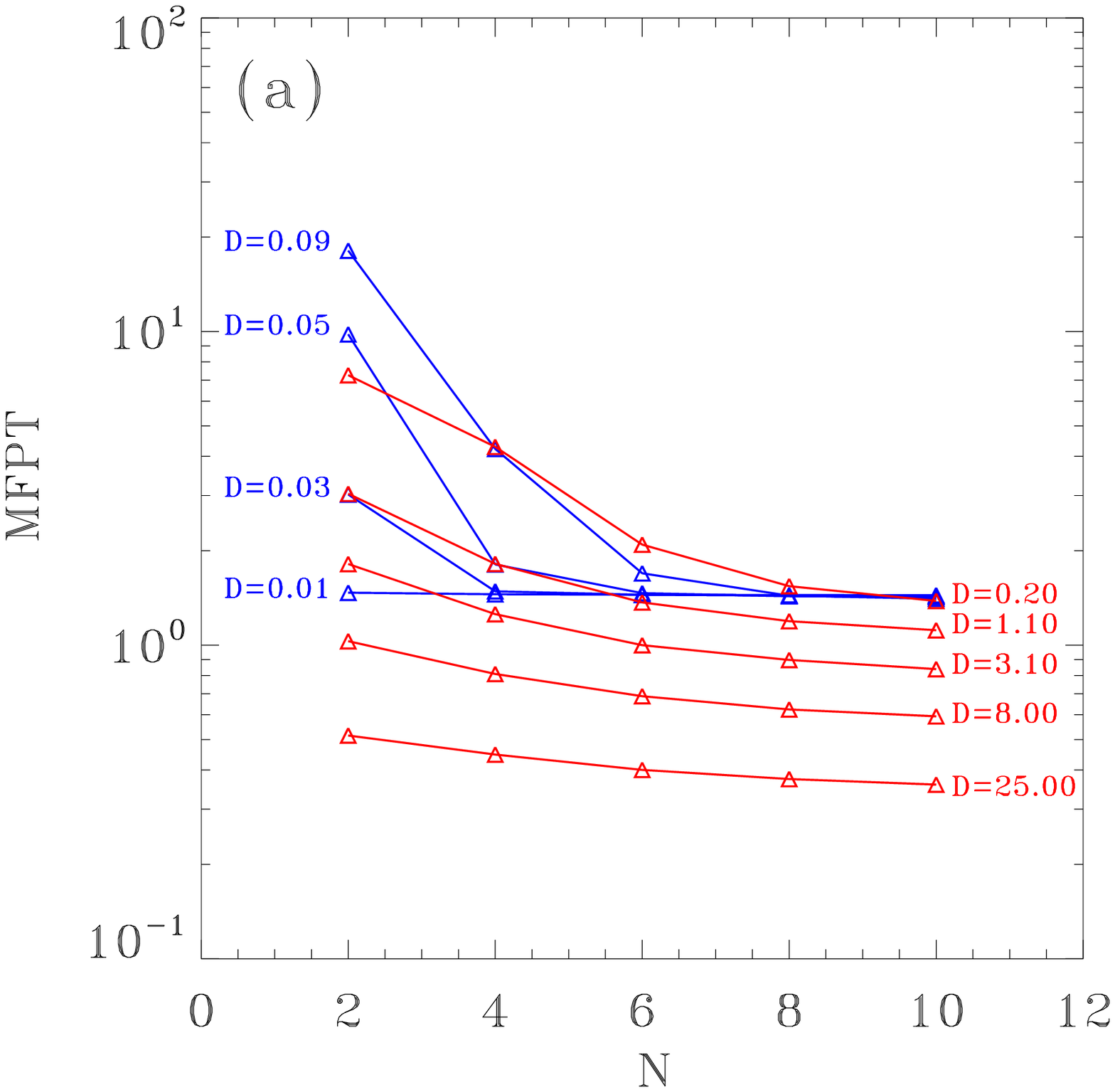}
\hspace{+0.5cm}
\includegraphics[width=8cm,height=7cm]{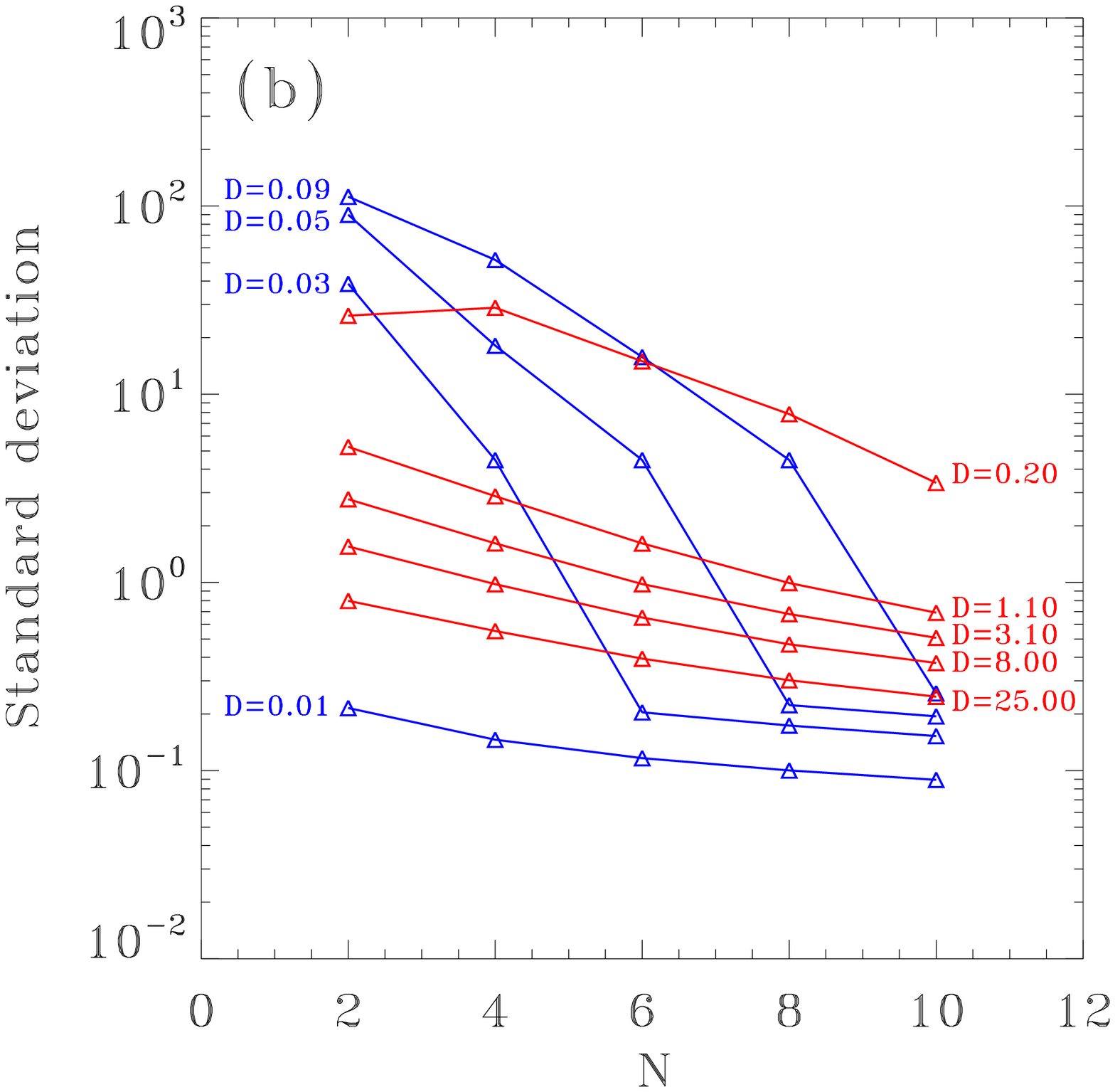}
\end{center}
\vskip-0.6cm \caption{\small (a): Dependence of the MFPT on the
number N of monomers at different noise intensities. Initial
conditions are the same as in Fig.~\ref{fig34}. Red lines are used
to discriminate the ascending trend of MFPTs with the decreasing of
the noise intensity with respect to the opposite behavior at
extremely low noise intesity (blue lines). (b): Standard deviation
of the passage time vs.~N for different levels of noise.}
\label{fig67}
\end{figure*}
The plot of the MFPT as a function of the polymer length
(Fig.~\ref{fig67}a) at different noise intensities shows that: (i)
the quasi-deterministic MFPT (noise level $D=0.01$) does not depend
on N; (ii) for $D>0.01$, the dependence of MFPT on the polymer
length shows a non-monotonic trend with the noise intensity:
differences of MFPT are very sensitive to the number of monomers
when the noise level is low, while they are reduced at higher noise
intensities. In Fig.~\ref{fig67}b it is interesting to note the
capture effect also for the polymer with N=4 and the low level of
noise $D=0.03$.

\section{\label{Conclus}Conclusions}
Experimental and theoretical studies on the translocation time,
taken by a chain molecule to cross a membrane, show a strong
dependence of this time with the length of the polymer. In this work
we propose a possible reason of this dependence by investigating the
stochastic dynamics of a polymer, with different lengths,
surmounting a potential barrier. After been trapped for a while, the
molecule leaves the top of the barrier exhibiting an enhancement of
the mean crossing time as a function of the temperature
fluctuations. This is a typical signature of the presence of a
metastable state in the system investigated. Our analysis shows a
clear effect of noise-enhanced stability on the mean first passage
time of short polymers for initial unstable conditions. Our
simulations demonstrate that longer polymers can travel the same
channel distance faster than shorter chains. This result is in
agreement with the experimental measurements of DNA mobility carried
out by Han \emph{et al.}~[1999], but a deeper analysis on the role
of initial conditions on the translocation time is necessary. Our
findings are also in accordance with the experimental results of
Meller \emph{et al.}~[2000], that is the longer translocation times
measured at lower temperatures, as we have found near the maxima of
MFPTs (see Fig.~\ref{fig34}a).

\section*{\label{sec:ack}Acknowledgements}
\vskip-0.48cm This work was supported by MIUR and CNISM. N.~P. wish
to thank Dr.~Martin Bear for a stimulating discussion and the
pertinent comments. A.~F. acknowledges the Marie Curie TOK grant
under the COCOS project (6th EU Framework Programme, contract No:
MTKD-CT-2004-517186).


\end{document}